\documentclass[runningheads]{llncs}
\usepackage{amsmath,amssymb}
\usepackage{multirow}
\usepackage[T1]{fontenc}
\usepackage{caption}
\usepackage{graphicx}

\usepackage{algorithm}
\usepackage{algpseudocode}
\usepackage{placeins}

\begin{document}
	\title{Differentially Private Datastore Generation for Retrieval-Augmented Inference}

\author{Abdelrahman Abouelenein
\inst{1,2}  
\and
Marwan Torki \inst{1}
}

\institute{Department of Computer and Systems Engineering
Alexandria University, Egypt 
% \and
\email{\{Abdelrahman, mtorki\}@alexu.edu.eg}
\and
Microsoft
}

\maketitle              % typeset the header of the contribution
\begin{abstract}
It is crucial for modern on-device AI systems that rely on retrieval-augmented inference to release and share
datastores without compromising individual privacy. This can be achieved using Differential Privacy (DP), which provides a formal guarantee that ensures individual contributions remain indistinguishable, even under adversarial analysis. In this paper, we introduce a hashing-based probability generation framework designed to enable the creation and release of differentially private datastores. Our approach employs locality-sensitive hashing (LSH) to efficiently partition high-dimensional data into buckets. We then add calibrated DP noise to the accumulated vote for each bucket, generating a probability distribution across classes. Our method is broadly applicable to any pipeline requiring secure key,value datastore creation and release. We conducted experiments on seven datasets with varying sample sizes and class counts, ranging from 2 to 14. At \(\varepsilon\)=5, our released DP datastore achieves strong privacy protection with only an average 2.6\% drop in accuracy. Finally, we benchmark DP datastore resilience to membership inference attacks, reducing attack accuracy to 53.60\%.

\end{abstract}
\section{Introduction}

Recently, there has been growing interest in pretraining large language models (LLMs) due to their coherent and superior natural language generation abilities across multiple NLP tasks, compared to smaller language models \cite{radford2019language,NEURIPS2020_1457c0d6}.
Nevertheless, pretrained LLMs still require fine-tuning, which becomes an essential tool to ground the model and align it to the user's preference.

Unfortunately, finetuning LLMs is challenging due to the size and sensitivity of these large models, which require considerable hyperparameter tuning and substantial resources to host and train them. Retrieval-augmented generation (RAG) has been proposed as an alternative to avoid fine-tuning altogether while achieving similar objectives in many cases.
The goal of the retriever is to fetch the most suitable information from the datastore according to the provided context and inject it into the LLM generation pipeline.

\begin{figure}
    \centering
    \includegraphics[width=90mm,scale=0.1]{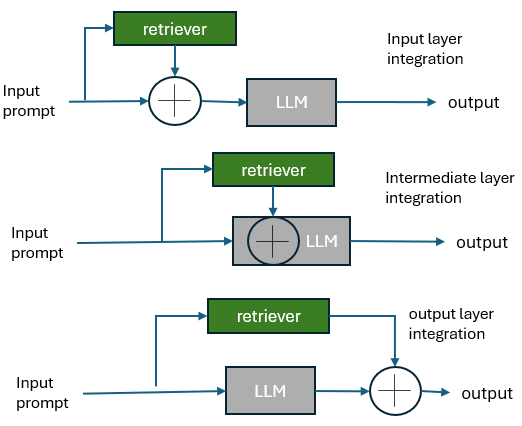}
    \caption{RAG Taxonomy based on the stage where the knowledge is being injected.}
    \label{fig:RAG_Taxonomy}
\end{figure}

Various frameworks exist for classifying RAG systems. As illustrated in Figure \ref{fig:RAG_Taxonomy}, one widely adopted taxonomy \cite{Cheng2025SurveyRAG} categorizes systems based on the specific stage at which the retriever incorporates external knowledge. Each RAG family offers distinct advantages; for example, input-layer integration retrieves exemplars into the context window to guide reasoning, while output-layer integration provides finer granularity by fusing retrieved distributions with the model's logits.

We categorize these key,value datastore dependent techniques under the umbrella of Retrieval-Augmented Inference (RAI). This term emphasizes that, whether the system performs logit-level fusion \cite{Khandelwal2020Generalization,Khandelwal2020NearestNM,efficientAndScalableKNNMT} or relies solely on the datastore for label assignment \cite{xu2023knn}, the final output is dynamically contingent upon a $k,v$ retrieval step performed at each inference step. The privacy of these pipelines remains an overlooked challenge.

Introduced by \cite{abadi2016}, differential privacy (DP) offers a formal way to reason about privacy risk. Intuitively, a mechanism is differentially private if its output is nearly the same whether or not any single individual record is included in the dataset, preventing an adversary from confidently inferring one person's participation. The parameter $\varepsilon$ quantifies this indistinguishability: smaller $\varepsilon$ implies stronger privacy, while larger $\varepsilon$ yields weaker privacy but typically better utility. Many practical systems use \,($\varepsilon,\delta$)\,--DP, where $\delta$ is a failure probability that allows the privacy guarantee to be violated on rare tail events. In practice, $\varepsilon$ is treated as a privacy budget: applying DP mechanisms multiple times increases the total privacy loss.
Some solutions \cite{igamberdiev-habernal-2023-dp,igamberdiev-etal-2022-dp} 
% \cite{igamberdiev-etal-2022-dp}
% \cite{igamberdiev-habernal-2023-dp}
attempted to rewrite the data with differential privacy (DP) guarantees, ensuring that the rewritten data is safe for public release. Unfortunately, the privacy budget $\varepsilon$ was sacrificed to rewrite the data while maintaining utility, yielding a minimal privacy guarantee. As stated by \cite{Ponomareva2023HowTD}, this is considered a tier 3 $\varepsilon$, making it susceptible to privacy attacks. 

On the other hand, \cite{tang2024privacypreserving} was one of the few works     to introduce DP to the in-context learning RAG framework. Although their proposed method provides good utility under strong DP guarantees, it is inefficient due to the inference cost of formulating differentially private shots.
While \cite{Panda2023PrivacyPreservingIL,tang2024privacypreserving} successfully introduced DP to the RAG framework, they fall short of bounding global privacy spending. In other words, each query increases the cumulative privacy cost, which could eventually lead to a privacy breach. To avoid this unbounded privacy-budget accumulation, we are motivated to generate a privacy-preserving datastore with a fixed privacy budget.

Locality-Sensitive Hashing (LSH) is often used as a heuristic for datastore privacy because it maps high-dimensional data to compact hash codes while approximately preserving similarity. Beyond privacy, hashing-based indexing can make retrieval substantially more efficient. Instead of scanning the full datastore, LSH uses hash tables to generate a small candidate set with simple operations such as bit comparisons and bucket lookups~\cite{ref_indyk_motwani}.
However, recent work shows that similarity-preserving hashes can still leak sensitive information and therefore should not be treated as privacy-preserving mechanisms on their own. In \cite{attack_on_google_minhash}, the authors show that MinHash/SimHash-based systems (e.g., Google’s FLoC and the MinHash Hierarchy) are vulnerable to inference and reconstruction attacks that operate solely on released hash values. We observe a similar leakage pattern in our setting: even with LSH, membership inference remains effective on non-private datastores, highlighting the need for DP.

Several recent work on DP approximate nearest‑neighbor rely on $\varepsilon$--$\delta$ differential privacy to achieve scalability and dimension independent utility. This is done via truncated noise mechanisms and advanced composition.
Notably, \cite{ref_andoni_dp_ann,ref_aumuller_range_counting,ref_dp_range_counting_hd} focus on answering fuzzy approximate counting queries in high‑dimensional spaces, where a nonzero $\delta$ is required to control tail events and cumulative privacy loss across many hash partitions.

Existing LSH‑based DP methods \cite{ref_andoni_dp_ann,ref_aumuller_range_counting,ref_dp_lsh} are designed to release scalar approximate counts at query time, and do not consider constructing a label‑level histogram suitable for downstream inference.
As a result, they cannot directly support multi‑class classification.
We extend this line of work from fuzzy approximate neighbor counting to histogram‑based vote aggregation. This enables multi‑class prediction while releasing the resulting datastore itself under pure $\varepsilon$‑DP guarantees, allowing unrestricted downstream use without additional privacy loss.

Motivated by these findings, we propose an approach that operates under pure $\varepsilon$‑differential privacy, eliminating any $\delta$‑dependent failure probability and providing strictly stronger privacy guarantees. This is an important distinction when releasing reusable data structures rather than answering individual queries.

To address these challenges, our contributions are:
{\renewcommand\labelitemi{$\ast$}
\begin{itemize}
    \item \textbf{DP SimHash Bucketing and Datastore Release.} We propose a hashing-based \(\varepsilon\)-DP framework that leverages Locality-Sensitive Hashing (LSH), specifically SimHash, to partition data into similarity-preserving buckets, and release a differentially private datastore that includes the hash functions and a noisy bucket-level class vote histogram.
    \item \textbf{Robust Retrieval at Query Time.} At query time, we hash the query into $T$ independent SimHash tables and retrieve the corresponding bucket aggregate from each table. Aggregating votes across tables mitigates hashing variance and improves recall.
    \item \textbf{Fixed Privacy Budget.} Our release is a one-shot mechanism that bounds the global privacy cost regardless of the number of downstream queries.
    \item \textbf{Integration with KNN-Prompting.} We demonstrate easy integration by applying our method to the KNN-Prompting \cite{xu2023knn} framework.
    \item \textbf{Empirical Study and Attack Evaluation.} We provide an extensive empirical study of the interaction between hashing parameters and DP parameters, and show resilience to membership inference attacks.
\end{itemize}}
\section{Background}
This section introduces the main concepts used throughout the paper. Retrieval-augmented inference describes how predictions can be made from a datastore at query time. LSH provides a way to group similar embeddings into reusable buckets. Differential privacy provides the formal guarantee used when releasing bucket-level class-count histograms.
The privacy guarantee relies on a bounded-contribution property: each record contributes to one class count in one bucket per hash table.

\subsection{Retrieval-Augmented Inference}  
Retrieval-augmented inference (RAI) uses a datastore at inference time: a query embedding retrieves nearby datastore entries, and their associated values are converted into a prediction distribution. Section 3 follows this template, but replaces exact neighbor retrieval with DP-noised bucket-level aggregates.

The core of the RAI framework lies in the generation of the non-parametric distribution, $p_{\text{kNN}}$. At each inference step $t$, the system encodes the current context into a query vector $\hat{h}_t$ and retrieves the $k$-nearest neighbors $\mathcal{N}_t$ from a datastore. The probability for a candidate token $\hat{y}_t$ is derived from the similarity of these neighbors to the query according to the following relation:

\begin{equation}
p_{\text{kNN}}(\hat{y}_t \mid x, y_{<t}) \propto 
\sum_{(h_i, y_i) \in \mathcal{N}_t} 
\mathbf{1}\!\left[\hat{y}_t = y_i\right] \,
\exp\!\left( -\frac{d(h_i, \hat{h}_t)}{\tau} \right),
\label{eq:knn-dist}
\end{equation}
where $d(\cdot, \cdot)$ is a task-appropriate distance or dissimilarity measure and $\tau$ is the temperature. This exponential weighting follows standard kNN-based retrieval formulations; the specific choice of $d$ depends on the retrieval setup.
Here, $d$ is a generic retrieval dissimilarity (e.g., Euclidean, squared Euclidean, or angular), and our DP release does not depend on this weighting form. While the generation of $p_{\text{kNN}}$ is consistent across RAI methods, its application depends on the specific architecture. In $k\text{NN-LM}$ and $k\text{NN-MT}$, this retrieved distribution is fused with the parametric language model distribution $p_{\text{LM}}$ via an interpolation coefficient $\lambda$, such that $p(\hat{y}) = \lambda p_{\text{kNN}}(\hat{y}) + (1 - \lambda) p_{\text{LM}}(\hat{y})$. In contrast, $k\text{NN-prompting}$, particularly in calibration-free settings, relies solely on the retrieved distribution $p_{\text{kNN}}$ to determine the final output.

\subsection{Locality-Sensitive Hashing for Approximate Nearest Neighbor}
Let $\mathcal{X}$ be a representation space equipped with a distance or dissimilarity measure $d(\cdot,\cdot)$. Given a dataset $D=\{x_1,\dots,x_n\}\subseteq\mathcal{X}$ and a query point $q\in\mathcal{X}$, the (exact) nearest neighbor problem returns
\begin{equation}
\operatorname{NN}(q;D)=\arg\min_{x\in D} d(q,x).
\end{equation}
Approximate nearest neighbor (ANN) relaxes this objective for efficiency, typically allowing a multiplicative or additive approximation.

In our method, this ANN notation is used only to motivate similarity-preserving bucketing. The actual implementation operates on normalized embedding vectors, and the privacy guarantee comes from the DP noise added to the released bucket histograms in Section 3.
Thus, neither the DP guarantee nor the datastore construction requires \(d\) to satisfy the metric axioms; in our experiments, \(\mathcal{X}\) is simply the normalized embedding space \(\mathbb{R}^m\), and \(d\) denotes the dissimilarity used for retrieval.

Hashing-based ANN methods construct an index through a hash family $\mathcal{H}$ that maps points to discrete keys (``buckets'') and supports sublinear candidate generation.
In the locality-sensitive hashing (LSH) framework, a hash family is designed so that nearby points collide with higher probability than far points.
One common formulation is:

\begin{definition}[$(r_1,r_2,p_1,p_2)$-sensitive hash family]
A family $\mathcal{H}$ over $\mathcal{X}$ is $(r_1,r_2,p_1,p_2)$-sensitive for distance $d$ if for any $x,y\in\mathcal{X}$:
\begin{align}
 d(x,y)\le r_1 &\Rightarrow \Pr_{h\sim\mathcal{H}}[h(x)=h(y)]\ge p_1,\\
 d(x,y)\ge r_2 &\Rightarrow \Pr_{h\sim\mathcal{H}}[h(x)=h(y)]\le p_2,
\end{align}
with $r_1<r_2$ and $p_1>p_2$.
\end{definition}

To improve selectivity, one typically concatenates $H$ independent hash bits into an $H$-bit code.

\subsubsection{SimHash (random hyperplane hashing)}
SimHash is an LSH technique that relies on cosine similarity / angular distance, making it suitable for approximate nearest neighbor search over normalized embeddings.
Let $x\in\mathbb{R}^m$ be a normalized embedding vector.
Draw $H$ random hyperplanes via vectors $r_1,\dots,r_H\in\mathbb{R}^m$ (e.g., i.i.d. standard normal or Rademacher entries).
The $i$-th SimHash bit is
\begin{equation}
g_i(x) = \mathbb{I}[\langle r_i, x\rangle \ge 0] \in \{0,1\},
\end{equation}
and the full code is represented as
\begin{equation}
g(x)=(g_1(x),\dots,g_H(x))\in\{0,1\}^H.
\end{equation}
All vectors $x$ in close proximity should have similar projections, thus ending up in the same bins.
These properties make SimHash attractive for fast candidate generation in high dimensions using cheap bit operations.

\subsection{Differential Privacy Mechanism}

For our release, the relevant privacy ingredients are pure $\varepsilon$-differential privacy under the add/remove neighboring relation, the Laplace mechanism, and composition across the released hash tables.

\begin{definition}[Differential Privacy]
A randomized mechanism $\mathcal{M}$ is $(\varepsilon)$-differentially private if for all neighboring datasets $\mathcal{D},\mathcal{D}'$ and all measurable outputs $S$.
\begin{equation}
\Pr[\mathcal{M}(\mathcal{D})\in S]\le e^{\varepsilon}\Pr[\mathcal{M}(\mathcal{D}')\in S].
\end{equation}
\end{definition}
We use the standard neighboring relation where datasets differ by one record in terms of addition or removal. In the definition above, \(S\) denotes any measurable subset of the output space of \(\mathcal{M}\). Thus, \(\mathcal{M}(D)\in S\) is the event that the randomized output of the mechanism falls in that subset.

In our datastore release, \(f(D)\) is vector-valued: it denotes the concatenation of all released bucket-level class-count histograms across the hash tables. Accordingly, \(\eta\) has the same dimension as \(f(D)\), with independent Laplace noise added to each entry. To ensure that a function \(f\) is \(\varepsilon\)-differentially private, we add noise drawn from a Laplace distribution.
\begin{equation}    
\mathcal{M}(D) = f(D) + \eta, \quad \eta \sim \mathrm{Laplace}(0, \beta)
\end{equation}
where $\beta$ is the Laplace scale parameter.
The scale parameter $\beta$ is:
\begin{equation}
\beta = \frac{\Delta f}{\varepsilon}
\end{equation}
The objective of the global sensitivity function \(\Delta f\) is to measure the maximum influence of a single record or sample when added or removed.For the Laplace mechanism, it is defined as
\begin{equation}
\label{sensitivity}
\Delta f = \max_{D, D' \text{ differ in one record}} \| f(D) - f(D') \|_1
\end{equation}

\subsubsection{Composition of Differential Privacy}

In most practical pipelines, multiple randomized mechanisms are applied during training, preprocessing, or release.
Composition theorems describe how privacy loss accumulates when combining such mechanisms \cite{ref_dwork}.
In Section 3, this is used to account for the fact that one record contributes to one bucket in each of the $T$ released hash tables.

\paragraph{Sequential composition.}
Let $\mathcal{M}_1,\dots,\mathcal{M}_m$ be mechanisms that are respectively $(\varepsilon_1)$, $\dots$, $(\varepsilon_m)$differentially private when applied to the same underlying dataset $D$.
Then the joint release
\begin{equation}
\mathcal{M}(D) = \big(\mathcal{M}_1(D),\dots,\mathcal{M}_m(D)\big)
\end{equation}
is $(\varepsilon_\text{tot})$-differentially private with
\begin{equation}
\varepsilon_\text{tot} = \sum_{j=1}^{m} \varepsilon_j.
\end{equation}
Intuitively, if a single record can influence the output of each mechanism, then privacy losses add up; $\varepsilon_\text{tot}$ is the overall privacy budget of the combined release.

\paragraph{Parallel composition.}
Let the dataset be partitioned into disjoint subsets $D = \biguplus_{j=1}^{m} D_j$.
If each mechanism $\mathcal{M}_j$ is $(\varepsilon_j)$-differentially private when applied only to its own subset $D_j$, then the combined release
\begin{equation}
\mathcal{M}(D)=\big(\mathcal{M}_1(D_1),\dots,\mathcal{M}_m(D_m)\big)
\end{equation}
is $(\varepsilon_\text{par})$-differentially private with
\begin{equation}
\varepsilon_\text{par} = \max_{j\in\{1,\dots,m\}} \varepsilon_j.
\end{equation}
This reflects the fact that any single record belongs to exactly one subset, and thus can affect at most one mechanism. Therefore $\varepsilon_\text{par}$ is the overall privacy budget of the combined release under parallel composition (it is not a sum because no record participates in multiple mechanisms). In practice, this overall budget is often reported simply as $\varepsilon$.

\section{DP Datastore Construction and Inference}

In this section, we present a release-and-reuse pipeline that privatizes a \(k,v\) datastore once and enables unlimited downstream queries without additional privacy cost.
Given normalized embeddings and labels \(\mathcal{D}_{\text{train}}=\{(a_i,y_i)\}_{i=1}^{N}\), we construct and release the DP datastore in three steps.

Our construction is agnostic to the origin of the embedding representation, including aggregate embedding descriptors~\cite{Torki2018DoCoV}, as long as each record maps to one datastore key and contributes one bounded vote per hash table:

\begin{enumerate}
\item \textbf{Bucketization (SimHash).} Hash each embedding into buckets using \(T\) independent SimHash tables with \(H\) random hyperplanes per table.
\item \textbf{Vote aggregation.} For each bucket, store a class-count histogram \(V_t[u]\in\mathbb{R}^C\) (one per table \(t\)).
\item \textbf{DP release.} Add calibrated Laplace noise to each stored bucket histogram and release the resulting datastore (hyperplanes + noisy votes).
\end{enumerate}

\begin{algorithm}[th!]
\caption{Differentially Private Datastore Creation }
\label{alg:simhash_build}
\begin{algorithmic}[1]
\Require Labeled embeddings $\mathcal{D}_{\text{train}}=\{(a_i,y_i)\}_{i=1}^{N}$, where $a_i \in \mathbb{R}^m$ and $y_i \in \{0,\dots,C-1\}$
\Require Random seed $s$, Privacy budget $\varepsilon$
\Require Number of tables $T$, number of hyperplanes $H$
\Ensure Released DP datastore $\mathcal{D}_{\text{DP}}=(R,V)$ where $R=\{R_t\}_{t=1}^{T}$ and $V=\{V_t\}_{t=1}^{T}$ (with DP-noised votes)

\State Sample (using seed $s$) $T$ independent sets of random hyperplanes:
\For{$t \gets 1$ to $T$}
    \State Draw $H$ random hyperplane normals $R_t = \{r_{t,1},\dots,r_{t,H}\}$ with $r_{t,h}\in\mathbb{R}^m$
    \State Normalize each $r_{t,h}$ to unit length
  \State Initialize vote table $V_t:\{0,\dots,2^H-1\}\rightarrow\mathbb{R}^C$ with default $\mathbf{0}$
\EndFor

\For{$i \gets 1$ to $N$} \Comment{insert a labeled embedding $(a_i,y_i)$}
    \For{$t \gets 1$ to $T$}
        \Comment{Compute $H$ bits via random projection}
        \For{$h \gets 1$ to $H$}
            \State $b_h \gets \mathbb{I}\left[\langle a_i, r_{t,h}\rangle > 0\right]$ \Comment{$b_h \in \{0,1\}$}
        \EndFor
        \State $u_i^{(t)} \gets \sum_{h=1}^{H} b_h \, 2^{h-1}$ \Comment{encode bits to bucket id, $u_i^{(t)} \in \{0,\dots,2^H-1\}$}
        \State $V_t[u_i^{(t)}][y_i] \gets V_t[u_i^{(t)}][y_i] + 1$ \Comment{update aggregated votes}
    \EndFor
\EndFor
\For{$t \gets 1$ to $T$}
\ForAll{bucket ids $u \in \{0,\dots,2^H-1\}$}
        \State Sample $\eta \sim \mathrm{Laplace}(0, \beta)^C$ \Comment{i.i.d. per class coordinate}
    \State $V_t[u] \gets V_t[u] + \eta$ \Comment{DP-fies bucket vote counts (including empty buckets)}
    \EndFor
    % \State Sample noise matrix $\eta_t \sim \mathrm{Laplace}(0,\beta)^{2^H \times C}$ \Comment{}
  % \State $V_t \gets V_t + \eta_t$ \Comment{Add noise to \emph{all} buckets in table $t$}
\EndFor

\State \Return $\mathcal{D}_{\text{DP}} \gets (R,V)$
\end{algorithmic}
\end{algorithm}

Algorithm \ref{alg:simhash_build} details datastore construction and the one-shot \(\varepsilon\)-DP release via Laplace noise.
For inference, Algorithm \ref{alg:simhash_infer} hashes a query \(q\) into each of the \(T\) tables and retrieves the corresponding bucket histograms.
The retrieved (DP-noised) vote vectors are summed across tables and the prediction is \(\hat{y}=\arg\max_{c} v_c\).

We choose SimHash (random hyperplane hashing) because our keys are normalized embedding vectors and retrieval is naturally governed by angular distance (equivalently, cosine similarity). Nearby vectors tend to share similar sign patterns, so Hamming distance between hash codes correlates with angular distance, enabling compact indexing.

To ensure our pipeline generates differentially private histograms per bucket, we apply the Laplace mechanism independently to each bucket-level class-count vector, in a manner similar to private knn \cite{dpknn}.

\paragraph{Sensitivity and noise calibration.}
Consider neighboring datasets that differ by the addition or removal of a single record. For each table $t$, that record hashes to exactly one bucket $u^{(t)}$ and increments exactly one class coordinate by 1. Therefore, across all $T$ tables, a single record can change at most $T$ histogram coordinates (one per table), each by magnitude 1.
Under the standard add/remove neighboring relation, the overall $\ell_1$ sensitivity of the released histogram vector is
\begin{equation}
\Delta f = T.
\end{equation}
Applying the Laplace mechanism to the full release then uses scale
\begin{equation}
\beta = \Delta f/\varepsilon = T/\varepsilon,
\end{equation}
which allows to release our aggregated vote under formal differential privacy. Finally, sampling and releasing the random hyperplanes $\{R_t\}$ does not consume privacy budget because they are generated independently of the training data.

\begin{algorithm}[t]
\caption{Inference with DP Datastore}
\label{alg:simhash_infer}
\begin{algorithmic}[1]
\Require Query vector $q \in \mathbb{R}^m$ 
\Require Released datastore $\mathcal{D}_{\text{DP}}=(R,V)$ where $R=\{R_t\}_{t=1}^{T}$, $V=\{V_t\}_{t=1}^{T}$, and each $R_t=\{r_{t,h}\}_{h=1}^{H}$
\Ensure Predicted label $\hat{y}$

\State Initialize aggregated votes $v \gets \mathbf{0}$ \Comment{same dimension as stored vote vectors}
\ForAll{tables $(R_t, V_t)$ in $(R,V)$}
    \State $u \gets g(q;R_t)$ \Comment{SimHash bucket id}
  \State $v \gets v + V_t[u]$ \Comment{$V_t$ is DP-noised for all $u \in \{0,\dots,2^H-1\}$}
\EndFor
\State $\hat{y} \gets \arg\max_{c} v_c$
\State \Return $\hat{y}$
\end{algorithmic}
\end{algorithm}

\FloatBarrier

\section{Experiments}
We transform the KNN-Prompting \(k,v\) datastore into a differentially private datastore using our algorithm and evaluate it on multiple text classification datasets. We use MR \cite{mr}, SUBJ \cite{pang-lee-2004-sentimental}, TREC \cite{trec}, CR \cite{cr}, AGNews, DBpedia \cite{dbpedia}, and MPQA \cite{Wiebe2005Annotating}, spanning 2 to 14 classes.

We evaluate on a fixed 1k-example test set per dataset except TREC uses its standard 500-example test set, and build the datastore from the corresponding training split. Across datasets, the datastore construction set ranges from 1.8k to 8.7k training samples (Table \ref{datasets_stats}) to stress-test privacy and utility in smaller-datastore regimes. We use Phi-4-Mini \cite{Microsoft2025Phi4Mini} as the backbone model. We fix the random seed to 42 for reproducibility.

We stratify both the training samples used for datastore construction and the test split to enforce equal class proportions for all datasets except TREC, CR, and MPQA. For these datasets, we use the standard full training split, which introduces class imbalance in datastore creation.  

Table \ref{tab:eps_accuracy} reports classification accuracy as a function of the privacy budget \(\varepsilon\). As expected, utility improves as \(\varepsilon\) increases (less noise). Across datasets, our released DP datastore achieves a favorable privacy--utility trade-off; for example, at \(\varepsilon=5\) we observe only a \(2.6\%\) average accuracy drop relative to non-private KNN-Prompting.

\begin{table}[ht!]
\centering
\small
\captionsetup{skip=5pt}
\setlength{\tabcolsep}{4pt}
\renewcommand{\arraystretch}{1.15}
\begin{tabular}{lrrr}
\hline
Dataset & Train & Test & \#Classes  \\
\hline
TREC    & 5452 & 500  & 6   \\
AGNews  & 8000 & 1000 & 4  \\
DBpedia & 8000 & 1000 & 14\\
CR      & 1775 & 1000 & 2  \\
MR      & 8662 & 1000 & 2 \\
SUBJ    & 8000 & 1000 & 2  \\
MPQA    & 8603 & 1000 & 2 \\
\hline
\end{tabular}
\caption{Dataset statistics. Test sets use 1k examples per dataset, except TREC.}
\label{datasets_stats}
\end{table}

\begin{table}[ht!]
\centering
\captionsetup{skip=5pt}
\setlength{\tabcolsep}{5pt}
\renewcommand{\arraystretch}{1.10}
\begin{tabular}{lcccccccc}
\hline
$\varepsilon$ & TREC & AGNews & CR & MR & SUBJ & DBpedia & MPQA & avg drop \\
\hline
KNN-Prompting & 94.4 & 88.2 & 89.2 & 89.7 & 93.3 & 98.8 & 85.4 & - \\
\hline
0.1 & 53.2 & 79.3  & 85.7  & 88.4 & 85.2 & 64.5 & 83.7 & 14.1\\ 
1 & 74.4 & 86.4 & 86.7 & 90.5 & 86.9 & 87.9 & 84.5 & 5.0 \\
3 & 84.2  & 87.7 & 88 & 90.6 & 86.9  & 93.6  & 84.5 & 3.4 \\
5 & 87.4 & 88.2 & 87.9 & 90.7 & 87.0 & 95.4 & 84.4 & 2.6 \\
7 & 89.6 & 88.2 &  88 & 90.8 & 87.1 & 95.9 & 84.5 & 2.1 \\
8 & 90.0 & 88.2 & 88 & 90.8 & 87.1 & 96.4 & 84.5 & 2.0 \\
\hline
\end{tabular}
\caption{Classification accuracy vs.\ total privacy budget $\varepsilon$ across datasets using 24 hyperplanes and 4 hash tables. Average drop is relative to non-private KNN-Prompting.}

\label{tab:eps_accuracy}
\end{table}

\subsection{Ablation Study}
We study the effect of the SimHash resolution (number of hyperplanes/bits) and the privacy budget on retrieval quality. Increasing the number of hyperplanes yields finer partitions (more buckets), which can improve nearest-neighbor selectivity but reduces the number of samples per bucket, making aggregated votes more sensitive to DP noise.

As shown in Table \ref{tab:trec_ablation}, increasing the number of hyperplanes yields finer partitions and can improve selectivity at higher \(\varepsilon\), but it reduces the number of samples per bucket, making aggregated votes more sensitive to DP noise.

Table \ref{tab:dbpedia_eps} shows the same trend on DBpedia; at 48 hyperplanes, accuracy drops from 91.0 at \(\varepsilon=8\) to 78.9 at \(\varepsilon=1\).

In contrast, using only 4 hyperplanes maps many samples to the same bucket, improving robustness to injected noise but reducing nearest-neighbor selectivity, since less similar samples are more likely to be grouped together.

Overall, Tables \ref{tab:trec_ablation} and \ref{tab:dbpedia_eps} lead to the same conclusion: there is a consistent trade-off between finer hashing (more hyperplanes) improving selectivity at higher \(\varepsilon\), versus coarser hashing being more robust under stronger privacy noise.

We additionally study the impact of datastore size (number of training examples included in the datastore). Increasing the datastore size improves results significantly on both TREC and DBpedia (Table \ref{tab:datastore_size}).

\begin{table}[ht!]
\centering
\captionsetup{skip=5pt}
\setlength{\tabcolsep}{6pt}
\renewcommand{\arraystretch}{1.15}
\begin{tabular}{lccccc}
\hline
$\varepsilon$ & 4 & 16 & 24 & 36 & 48 \\
\hline
1.0 & 61.6 & 77.6 & 74.4 & 66.6 & 66.8 \\
3.0 & 65.8 & 80.0 & 84.2 & 76.4 & 73.4 \\
5.0 & 64.4 & 80.6 & 87.4 & 80.2 & 76.8 \\
8.0 & 64.4 & 80.6 & 90 & 83.2 & 80.2 \\
\hline
\end{tabular}
\caption{TREC dataset accuracy for different numbers of hyperplanes and privacy budgets ($\varepsilon$).}
\label{tab:trec_ablation}
\end{table}

\begin{table}[ht!]
  \centering
\captionsetup{skip=5pt}
  \setlength{\tabcolsep}{6pt}
\renewcommand{\arraystretch}{1.15}
  \begin{tabular}{lcccccc}
    \hline
    $\varepsilon$ & 4 & 8  & 16 & 24 & 32 & 48 \\
    \hline
    1.0 & 84.9 & 91.7 & 92.8 & 87.9 & 84.7 & 78.9 \\
3.0 & 84.9 & 92.6 & 95.3 & 92.5 & 90.9 & 86.3 \\
5.0 & 84.9 & 92.6 & 96.5 & 94.5 & 93.3 & 89.3 \\
8.0 & 84.9 & 92.7 & 97.3 & 96.1 & 94.3 & 91.0 \\
    \hline
  \end{tabular}
  \caption{DBpedia dataset accuracy for different numbers of hyperplanes and privacy budgets ($\varepsilon$).}
  
  \label{tab:dbpedia_eps}
\end{table}

\begin{table}[ht!]
\centering
\small
\captionsetup{skip=5pt}

\setlength{\tabcolsep}{6pt}
\renewcommand{\arraystretch}{1.15}
\begin{tabular}{llccc}
\hline
 Datastore & 100 & 400 & 800 & full \\
\hline
 Ours & 71.4 & 72.4 & 76.2 & 90.0 \\
 KNN-Prompting & 79.2 & 89.2 & 91.2 & 94.4 \\
\hline
\end{tabular}
\caption{Ablation on datastore size (number of training examples included in the datastore): accuracy (\%).}
\label{tab:datastore_size}
\end{table}

Table \ref{tab:datastore_size} shows that increasing the number of samples used to construct the datastore consistently improves accuracy, and it also reduces the gap between the DP datastore and the original (non-private) one. On TREC, scaling from a 100-example datastore to using the full training set increases our accuracy from 71.4\% to 90.0\%. Over the same range, the gap to KNN-Prompting shrinks from 7.8 points at 100 examples (79.2\% vs. 71.4\%) to 4.4 points with the full datastore (94.4\% vs. 90.0\%). This indicating that larger datastores help both utility and DP robustness while closing the gap with original datastore.

\subsection{Membership Inference Attack}

Inspired by \cite{Shokri2016MembershipIA}, we evaluate resilience to membership inference attacks (MIA) on the released datastore. The attacker is given access to the released DP datastore (hyperplanes and noisy bucket histograms) and can issue unlimited queries to observe responses from our datastore. We build balanced evaluation sets with matching class distributions between members and non-members.

In our setting, the DP datastore is built from 8k DBpedia stratified training samples. The attacker has access to 50\% of training samples use; we use 4k member queries and 4k non-member queries sampled from Test set. We use the following query-response features: (1) per-class votes, (2) maximum vote, (3) vote confidence (max/total), (4) vote entropy, and (5) bucket hit rate (fraction of tables with non-empty buckets).

Without DP (\(\varepsilon=\infty\)), the attacker achieves 70.22\% accuracy even with hashing. As \(\varepsilon\) decreases, added noise reduces distinguishability and drives attack accuracy toward random guessing.Because our member and non-member evaluation sets are balanced (4k/4k), random guessing yields 50\% attack accuracy. We therefore also report the attack advantage, defined as $\mathrm{Adv} = \mathrm{Acc} - 0.5$, to highlight how far the attacker is above random chance.

\paragraph{Empty-bucket leakage and fairness.}
In non-private deployments, it is common to store only non-empty buckets. In that setting, if a query hashes to an absent (empty) bucket, the attacker can trivially infer non-membership, creating an unfair structural leakage channel. Our DP release avoids this by adding calibrated noise to all buckets in each hash table (including empty buckets), so every query returns a (possibly noise-only) vote vector and bucket existence is not an informative signal.

\begin{table}[ht!]
  \centering
\captionsetup{skip=5pt}
  \setlength{\tabcolsep}{6pt}
\renewcommand{\arraystretch}{1.15}
  \label{tab:mia}
  \begin{tabular}{|c|c|c|}
\hline
$\varepsilon$ & Attack accuracy & Advantage \\ \hline
 \(\infty\) & 70.22 & +20.22 \\ \hline
 8.0 & 60.02 & +10.02 \\ \hline
 7.0 & 58.80 & +8.80 \\ \hline
 5.0 & 57.20 & +7.20 \\ \hline
 3.0 & 53.60 & +3.60 \\ \hline

\end{tabular}
\caption{Membership inference attack on DBpedia.}
\end{table}

\FloatBarrier

\section{Conclusion and Future Work}

We introduced a release-and-reuse framework for retrieval-augmented inference that enables publishing a \emph{reusable} key--value datastore with a fixed privacy budget under pure $\varepsilon$-differential privacy. The core idea is to avoid releasing sensitive embeddings directly and instead privatize \emph{bucket-level} class-vote aggregates produced by similarity-preserving SimHash bucketing. This produces a compact datastore (random hyperplanes + DP-noised vote tables) that can be shared and queried without additional privacy spending. Across seven text-classification datasets (2 to 14 classes), the released DP datastore preserves utility well: at $\varepsilon=5$ we observe only a 2.6\% average accuracy drop relative to non-private KNN-Prompting. We also evaluate membership inference on the released datastore and find that DP noise substantially reduces attack success, pushing the attacker close to random guessing as $\varepsilon$ decreases.

Several directions can further strengthen and broaden the approach. (1) \textbf{Improved similarity under hashing:} incorporating weighted random projections could better align bucketing with task-specific embedding geometry, improving retrieval selectivity at a fixed privacy level. (2) \textbf{Beyond classification:} extending bucket-level aggregation from label histograms to token-level distributions would enable applying the same one-shot DP datastore release to kNN-LM and kNN-MT, where large vocabularies and interpolation with $p_{\text{LM}}$ introduce new efficiency and calibration challenges. (3) \textbf{Reducing hyperparameter tuning:} our current pipeline requires tuning hashing and inference hyperparameters (e.g., $H$, $T$ and related retrieval settings) to balance selectivity and DP noise; we plan to simplify this via more robust defaults and data-driven parameter selection.


\begin{thebibliography}{99}


\bibitem{Shokri2016MembershipIA}
Shokri, R., Stronati, M., Song, C., Shmatikov, V.:
Membership Inference Attacks Against Machine Learning Models.
In: \emph{IEEE Symposium on Security and Privacy (S\&P)}, pp. 3--18 (2017)

\bibitem{tang2024privacypreserving}
Tang, X., Shin, R., Inan, H.A., Manoel, A., Mireshghallah, F., Lin, Z., Gopi, S., et al.:
Privacy-Preserving In-Context Learning with Differentially Private Few-Shot Generation.
In: \emph{ICLR} (2024)

\bibitem{Khandelwal2020Generalization}
Khandelwal, U., Levy, O., Jurafsky, D., Zettlemoyer, L., Lewis, M.:
Generalization through Memorization: Nearest Neighbor Language Models.
In: \emph{ICLR} (2020)

\bibitem{Panda2023PrivacyPreservingIL}
Wu, T., Panda, A., Wang, J.T., Mittal, P.:
Privacy-Preserving In-Context Learning for Large Language Models.
In: \emph{ICLR} (2024)

\bibitem{dbpedia}
Zhang, X., Zhao, J., LeCun, Y.:
Character-level Convolutional Networks for Text Classification.
In: \emph{NeurIPS}, vol. 28 (2015)

\bibitem{cr}
Hu, M., Liu, B.:
Mining and Summarizing Customer Reviews.
In: \emph{Proc. ACM SIGKDD}, pp. 168--177 (2004)

\bibitem{trec}
Voorhees, E.M., Tice, D.M.:
Building a Question Answering Test Collection.
In: \emph{Proc. ACM SIGIR}, pp. 200--207 (2000)

\bibitem{mr}
Pang, B., Lee, L.:
Seeing Stars: Exploiting Class Relationships for Sentiment Categorization with Respect to Rating Scales.
In: \emph{Proc. ACL}, pp. 115--124 (2005)

\bibitem{pang-lee-2004-sentimental}
Pang, B., Lee, L.:
A Sentimental Education: Sentiment Analysis Using Subjectivity Summarization Based on Minimum Cuts.
In: \emph{Proc. ACL}, pp. 271--278 (2004)

\bibitem{dpknn}
Zhu, Y., Yu, X., Chandraker, M., Wang, Y.-X.:
Private-kNN: Practical Differential Privacy for Computer Vision.
In: \emph{CVPR}, pp. 11851--11859 (2020)

\bibitem{xu2023knn}
Xu, B., Xie, Y., Gu, Z., Zhu, K., Peng, S.:
kNN Prompting: Beyond-Context Learning with Calibration-Free Nearest Neighbor Inference.
In: \emph{ICLR} (2023)

\bibitem{efficientAndScalableKNNMT}
Dai, Y., Zhang, Z., Liu, Q., Cui, Q., Li, W., Du, Y., Xu, T.:
Simple and Scalable Nearest Neighbor Machine Translation.
arXiv preprint arXiv:2302.12188 (2023)

\bibitem{Khandelwal2020NearestNM}
Khandelwal, U., Fan, A., Jurafsky, D., Zettlemoyer, L., Lewis, M.:
Nearest Neighbor Machine Translation.
In: \emph{ICLR} (2021)

\bibitem{Wiebe2005Annotating}
Wiebe, J., Wilson, T., Cardie, C.:
Annotating Expressions of Opinions and Emotions in Language.
\emph{Language Resources and Evaluation}, 39(2--3), pp. 165--210 (2005)

\bibitem{NEURIPS2020_1457c0d6}
Brown, T., Mann, B., Ryder, N., Subbiah, M., Kaplan, J.D., Dhariwal, P., Neelakantan, A., et al.:
Language Models are Few-Shot Learners.
In: \emph{NeurIPS}, pp. 1877--1901 (2020)

\bibitem{radford2019language}
Radford, A., Wu, J., Child, R., Luan, D., Amodei, D., Sutskever, I.:
Language Models are Unsupervised Multitask Learners.
OpenAI Technical Report (2019)

\bibitem{igamberdiev-habernal-2023-dp}
Igamberdiev, T., Habernal, I.:
DP-BART for Privatized Text Rewriting under Local Differential Privacy.
In: \emph{Findings of the Association for Computational Linguistics}, pp. 13914--13934 (2023)

\bibitem{Ponomareva2023HowTD}
Ponomareva, N., Kurakin, A., Chien, S., Thakurta, A., Matthews, P.:
How to DP-fy ML: A Practical Guide to Machine Learning with Differential Privacy.
arXiv preprint arXiv:2303.00654 (2023)

\bibitem{Cheng2025SurveyRAG}
Cheng, M., Zhao, W.X., Zhang, J., Wen, J.-R.:
A Survey on Knowledge-Oriented Retrieval-Augmented Generation.
arXiv preprint arXiv:2503.10677 (2025)

\bibitem{igamberdiev-etal-2022-dp}
Igamberdiev, T., Habernal, I.:
DP-Rewrite: Towards Reproducibility and Transparency in Differentially Private Text Rewriting.
In: \emph{COLING}, pp. 2927--2933 (2022)

\bibitem{abadi2016}
Abadi, M., Chu, A., Goodfellow, I., McMahan, H.B., Mironov, I., Talwar, K., Zhang, L.:
Deep Learning with Differential Privacy.
In: \emph{ACM CCS} (2016)

\bibitem{ref_indyk_motwani}
Indyk, P., Motwani, R.:
Approximate Nearest Neighbors: Toward Removing the Curse of Dimensionality.
In: \emph{STOC}, pp. 604--613 (1998)

\bibitem{ref_dwork}
Dwork, C., Roth, A.:
The Algorithmic Foundations of Differential Privacy.
\emph{Foundations and Trends in Theoretical Computer Science}, 9(3--4), pp. 211--407 (2014)

\bibitem{ref_andoni_dp_ann}
Andoni, A., Dadush, D., Klein, N., Liu, K., Zhang, L.:
Differentially Private Approximate Near Neighbor Counting in High Dimensions.
In: \emph{NeurIPS} (2023)

\bibitem{ref_aumuller_range_counting}
Aum{\"u}ller, M., Gollapudi, S., Pagh, R., Silvestri, F.:
Differentially Private High-Dimensional Approximate Range Counting, Revisited.
In: \emph{Symposium on Foundations of Responsible Computing} (2025)

\bibitem{ref_dp_lsh}
Kenthapadi, K., Korolova, A., Mironov, I., Mishra, N.:
Differential Privacy with Locality-Sensitive Hashing.
In: \emph{ACM SIGKDD} (2012)

\bibitem{ref_dp_range_counting_hd}
Chan, T.-H.H., Li, M., Shi, E., Xu, W.:
Differentially Private Approximate Range Counting in High Dimensions.
In: \emph{ICALP} (2011)

\bibitem{Microsoft2025Phi4Mini}
Abouelenin, A., Ashfaq, A., Atkinson, A., Awadalla, H., Bach, N., et al.:
Phi-4-Mini Technical Report: Compact yet Powerful Multimodal Language Models via Mixture-of-LoRAs.
arXiv preprint arXiv:2503.01743 (2025)

\bibitem{attack_on_google_minhash}
Turati, F., Kubicek, K., Cotrini, C. and Basin, D.: Locality-Sensitive Hashing Does Not Guarantee Privacy! Attacks on Google's FLoC and the MinHash Hierarchy System. In: \emph{Proceedings on Privacy Enhancing Technologies}, 2023(4), pp. 117--131 (2023)


\bibitem{Torki2018DoCoV}
Torki, M.: A Document Descriptor using Covariance of Word Vectors.
In: \emph{Proceedings of the 56th Annual Meeting of the Association for Computational Linguistics (Short Papers)}, pp. 527--532 (2018)


\end{thebibliography}
\end{document}